\documentclass[reprint,aps,prl,twocolumn,showpacs,amsmath,amssymb]{revtex4}
\usepackage[dvipdfmx]{graphicx}
\usepackage[dvipdfmx]{color}
\usepackage{url}
\usepackage{dcolumn}
\usepackage{bm}
\usepackage{xspace}

\begin{document}


\title{Evidence of competing interactions within hole Fermi-surface sheets in the vicinity of KFe$_2$As$_2$}



\author{Y.~Ota$^{1}$}
\author{K.~Okazaki$^{1}$}
\altaffiliation{Present address: Department of Physics, University of Tokyo, Tokyo 113-0033, Japan}
\author{Y.~Kotani$^{1}$}
\altaffiliation{Present address: Japan Synchrotron Radiation Research Institute (JASRI/SPring-8), Sayo, Hyogo 679-5198, Japan}
\author{T.~Shimojima$^{1}$}
\altaffiliation{Present address: Department of Applied Physics, University of Tokyo, Tokyo 113-8656, Japan}
\author{W.~Malaeb$^{1}$}
\author{S.~Watanabe$^{2}$}
\author{C.-T.~Chen$^{3}$}
\author{K.~Kihou$^{4}$}
\author{C.~H.~Lee$^{4}$}
\author{A.~Iyo$^{4}$}
\author{H.~Eisaki$^{4}$}
\author{T.~Saito$^{5}$}
\author{H.~Fukazawa$^{5}$}
\author{Y.~Kohori$^{5}$}
\author{S.~Shin$^{1,6,7}$}
\affiliation{
$^{1}$Institute for Solid State Physics (ISSP), University of Tokyo, Kashiwa, Chiba 277-8581, Japan\\
$^{2}$Research Institute for Science and Technology, Tokyo University of Science, Chiba 278-8510, Japan\\
$^{3}$Beijing Center for Crystal R\&D, Chinese Academy of Science (CAS), Zhongguancun, Beijing 100190, China\\
$^{4}$National Institute of Advanced Industrial Science and Technology (AIST), Tsukuba, Ibaraki 305-8568, Japan\\
$^{5}$Department of Physics, Chiba University, Chiba 263-8522, Japan\\
$^{6}$CREST, JST, Chiyoda-ku, Tokyo 102-0075, Japan\\
$^{7}$RIKEN SPring-8 Center, Sayo-gun, Hyogo 679-5148, Japan
}



\date{\today}

\begin{abstract}
We have investigated the superconducting(SC)-gap anisotropy of Ba$_{0.07}$K$_{0.93}$Fe$_2$As$_2$, Ba$_{0.12}$K$_{0.88}$Fe$_2$As$_2$, and Ba$_{0.24}$K$_{0.76}$Fe$_2$As$_2$ using laser-based angle-resolved photoemission spectroscopy. Whereas in the end material K122, the inner and middle Fermi surfaces (FSs) have substantially anisotropic SC gaps, the outer FS has an almost zero gap, and the middle FS has octet-line nodes, in the 7\% Ba-doped sample, the SC gap of the inner FS becomes almost isotropic, the middle has a still anisotropic gap and nodes, and the outer has a finite gap and nodes. In addition, in the 12\% and 24\% Ba-doped samples, the SC gap of the middle FS becomes almost isotropic. We attribute this doping dependence to the existence of competing pairing interactions in this doping region.
\end{abstract}

\pacs{74.25.Jb, 74.70.Xa, 71.18.+y, 79.60.-i}

\maketitle



%


Superconductivity of iron-pnictides are induced by carrier doping in most cases. Hence, doping dependence of superconducting (SC) order parameter can reveal important character in the paring interactions. Angle-resolved photoemission spectroscopy (ARPES) is a powerful tool for the direct observation of the SC gap, it has been extensively utilized to the iron-based superconductors~\cite{Ding2008EPL} since their discovery.~\cite{Kamihara2008JACS, Rotter2008PRL} At the early stage, most of the ARPES studies reported the almost isotropic SC gap, although some thermodynamic properties have suggested anisotropic or nodal SC gaps.~\cite{Hashimoto2010PRB,Zeng2010NC,Yamashita2011PRB} Recently, several ARPES studies reported the anisotropic~\cite{Umezawa2012PRL,Borisenko2012Symmetry,Okazaki2012PRL} and nodal~\cite{Zhang2012NP,Okazaki2012Science} SC gaps. Among the iron-based superconductors which have anisotropic and/or nodal SC gaps, Ba$_{1-x}$K$_{x}$Fe$_2$As$_2$ (BaK122) is particularly unique. K atom can be completely substituted for Ba site of the parent compound BaFe$_2$As$_2$, and the end member KFe$_2$As$_2$ (K122) still shows a superconductivity with a critical temperature ($T_c$) = 3.4 K, whereas 40 \% substitution is the optimal doping and the highest $T_c$ of $\sim$ 38 K can be obtained. The optimally-doped (OP) member has hole FS sheets around the Brillouin-zone (BZ) center ($\Gamma$ point) and electron sheets around the BZ corner ($X$ point), and the SC gaps of these FS sheets are fully gapped (no node in the SC gap).~\cite{Ding2008EPL,Nakayama2009EPL} On the other hand, K122 had been suggested to have SC-gap nodes from several studies,~\cite{Fukazawa2009JPSJ,Dong2010PRL,Hashimoto2010PRBa} and its SC-gap symmetry had been proposed as a nodal $s$-wave~\cite{Kawano-Furukawa2011PRB,Suzuki2011PRB} or $d$-wave~\cite{Reid2012PRL,Thomale2011PRLa} both experimentally and theoretically. 

Laser-ARPES studies revealed that in the OP BaK122 the SC-gap anisotropies and FS-sheet dependence are weak for the three hole FSs around the BZ center,~\cite{Shimojima2011Science} and that in K122 they are strong and the SC-gap nodes exist in the middle FS at the FS angle $\varphi$ $\sim$ $\pm$ 5$^\circ$.~\cite{Okazaki2012Science} In addition, it has been found that the SC-gap size of the outer FS becomes abruptly reduced and vanishingly small for the overdoped (OD) region at the doping level where the electron FS at the BZ corner disappears ($x$ $\sim$ 0.6).~\cite{Malaeb2012PRB} 

Because the SC-gap anisotropies of K122 are significantly different from those of OP and OD BaK122 typically for the inner and middle FS sheets, it is very interesting to investigate how the SC-gap anisotropies of three hole FS sheets around BZ center change with Ba doping from K122. We have measured SC-gap sizes of 7 \% Ba-doped sample Ba$_{0.07}$K$_{0.93}$Fe$_2$As$_2$ (BaK0.93, $T_c$ $\sim$ 7 K), 12 \% Ba-doped sample Ba$_{0.12}$K$_{0.88}$Fe$_2$As$_2$ (BaK0.88, $T_c$ $\sim$ 13 K) and 24 \% Ba-doped sample Ba$_{0.24}$K$_{0.76}$Fe$_2$As$_2$ (BaK0.76, $T_c$ $\sim$ 17 K) using a low-temperature high-resolution laser-ARPES apparatus that achieves the highest energy resolution of 70 $\mu$eV and the lowest temperature of 1.5 K.~\cite{Okazaki2012Science},
and find that the SC-gap anisotropies of three hole FS sheets drastically change with a small amount of Ba doping. In addition, although the SC-gap sizes of the outer FS were vanishingly small for K122 and Ba$_{0.3}$K$_{0.7}$Fe$_2$As$_2$ (BaK0.7),~\cite{Malaeb2012PRB} those of BaK0.93, BaK0.88, and BaK0.76 show strong anisotropies and nodes exist for these compounds. On the other hand, for the middle FS, whereas SC-gap nodes exist for BaK0.93, the anisotropy is strongly suppressed for BaK0.88 and BaK0.76. We attribute this drastic doping dependence to the existence of the competing interactions as Maiti {\it et al.} proposed that the intraband and interband interactions are nearly degenerate.~\cite{Maiti2012PRB}

High quality single crystals were grown by self-flux method with KAs flux as described in Ref.~\onlinecite{Kihou2010JPSJ}. Clean surfaces were obtained by cleaving the sample {\it in situ} under ultra-high vacuum better than $5 \times 10^{-11}$ Torr. ARPES data were collected using the laser-ARPES apparatus at ISSP with the 6.994 eV, 6th harmonics of Nd:YVO$_4$ quasi-continuous wave (q-CW, repetition rate = 240 MHz) laser and VG-Scienta HR8000 electron analyzer as described in Ref.~\onlinecite{Okazaki2012Science}. The overall energy resolution was set to $\sim$ 1.2 meV for the SC-gap measurements and $\sim$ 4 meV for the FS and $E$-$k$ maps, and the angular resolution was 0.1 deg. Polarization of the incident excitation laser was adjusted using half-wave ($\lambda$/2) and quarter-wave ($\lambda$/4) plates. The superconducting transition temperatures were determined by the magnetization measurements before and after the ARPES measurements. The temperature dependence of the ARPES spectra were consistent with the magnetization measurements (see Fig. S1).


\begin{figure}[t]
\begin{center}
\includegraphics[width=8.5cm]{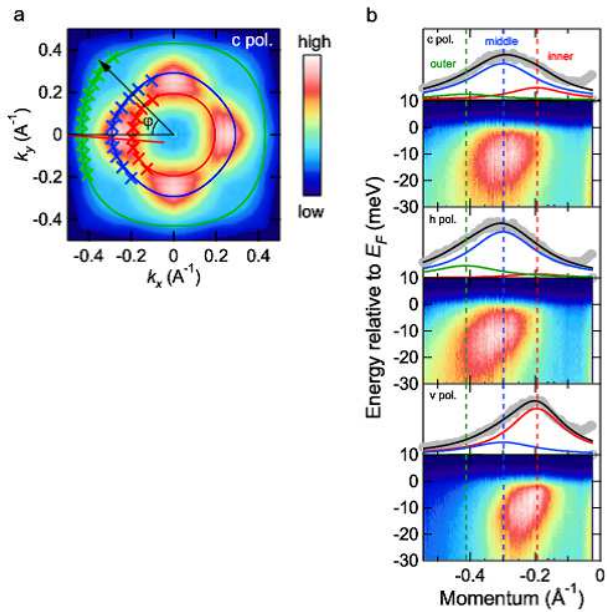}
\caption{FS map and MDCs at $E_F$ for BaK0.93. (a) FS map measured with $c$-polarized light. The integration energy window is $\pm$5 meV from $E_F$. The image is symmetrized assuming a crystal tetragonal symmetry. Red, blue, and green markers indicate the positions of $k_F$ for inner-, middle-, and outer-hole FSs, respectively. Definition of the FS angle $\varphi$ is shown. (b) MDCs at $E_F$ and $E$-$k$ maps obtained with $c$-, $h$-, and $v$-polarized light for $\varphi$ $\sim$  0$^\circ$ as indicated by the red line in (a). In these measurements, we used the experimental configurations similar to the previous study of K122.~\cite{Okazaki2012Science} }
\end{center}
\end{figure}

First, we determined the positions of Fermi momentum ($k_F$) for three (inner, middle, and outer) hole FSs around the BZ center. The FS map of BaK0.93 is shown in Fig. 1a. This FS map was measured with circularly($c$)-polarized light and symmetrized by assuming the tetragonal crystal structure (see Fig. S2 for raw data and other polarizations). The momentum distribution curves (MDCs) at $E_F$ and $E$-$k$ maps are shown in Fig. 1b, measured with the $c$-, horizontally($h$)- and vertically($v$)-polarized light~\cite{Okazaki2012Science} for $\varphi$ $\sim$ 0$^\circ$. The $k_F$ positions of three hole FSs were determined by fitting to the three Lorentzians as in the case of K122, and are indicated by the symbols on the FS map in Fig. 1a. The polarization dependence of these MDCs is similar to that of K122, and thus the dominant orbital character of each FS sheet also should be similar to that of K122, i.e., the inner, middle, and outer FSs are dominantly contributed from $xz/yz$, $xz/yz+z^2$, and $x^2-y^2$ orbitals, respectively. We fitted the $k_F$ positions to the following model function,
\begin{equation}
k_F(\varphi)=k_{F0}[1+A\cos(4\varphi)].
\end{equation}
From this fitting, we got the cross-sectional area of each FS sheet. With the same procedures, we determined the $k_F$ positions and FS areas for BaK0.88 and BaK0.76 (Figs. S3-S5). The results are shown in Table I. The doping dependence of the FS areas is consistent with the nominal change of the Fe valence.

\begin{figure*}[htbp]
\begin{center}
\includegraphics[width=12cm]{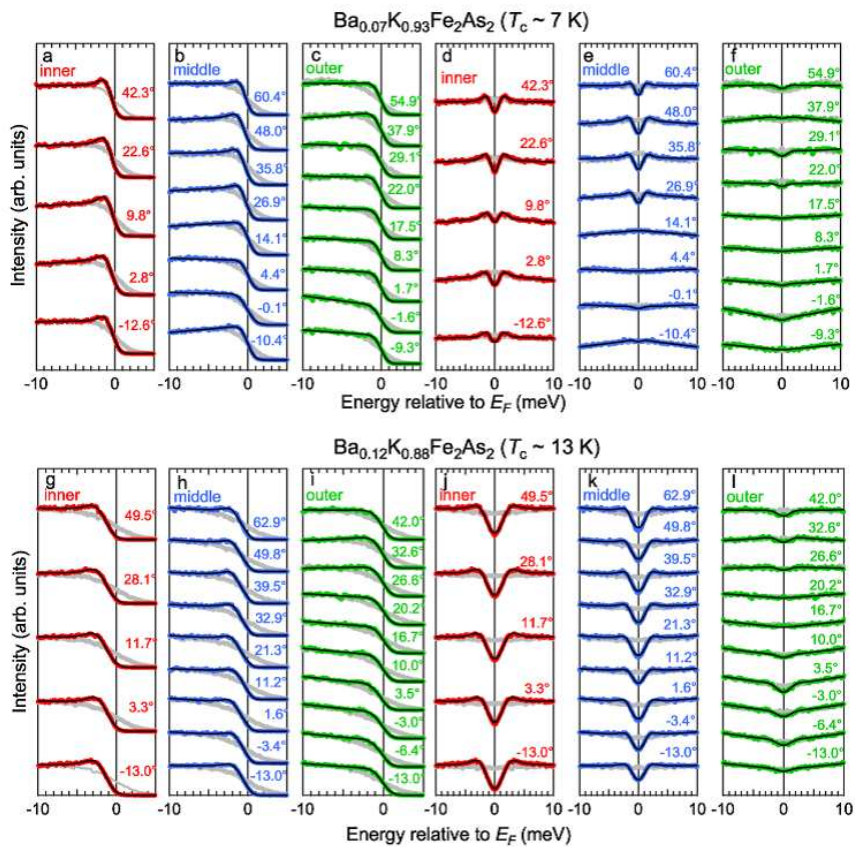}
\caption{EDCs and symmetrized EDCs at $k_F$ measured below $T_c$ and above $T_c$. (a)-(c) EDCs at $k_F$ of inner (a), middle (b), and outer FSs (c), respectively, for BaK0.93 measured at 1.5 K and 10 K ($T_c$ $\sim $ 7 K) for various FS angles as shown in each panel. (d)-(f), Symmetrized EDCs of inner (d), middle (e), and outer FSs (f), respectively. (g)-(i), EDC at $k_F$ of inner (g), middle (h), and outer FSs (i), respectively, for BaK0.88 measured at 1.5 K and 16 K ($T_c$ $\sim $ 13 K). (j)-(l), Symmetrized EDCs of inner (j), middle (k), and outer FSs (l), respectively. Gray markers indicate the EDCs above $T_c$ and the solid lines indicate the fitting functions using a BCS spectral function. The results for BaK0.76 are shown in Fig. S6.}
\end{center}
\end{figure*}

\begin{table} [t]
\begin{center}
\caption{Fitting parameters for $k_F$ positions using equation (1) and cross-sectional FS area. The areas are expressed as a percentage of the area of the 2D BZ.}
{
\begin{tabular}{|c|c|c|c|}
\multicolumn{4}{c}{inner} \\ \hline
~  $x$ ~ & $k_{F0}$ & A &area \\ \hline
1 & 0.22 & 0.02 & 5.9 \\ \hline 
0.93 & 0.20 & -0.02 & 4.7 \\ \hline 
0.88 & 0.18 & -0.06 & 3.9 \\ \hline 
0.76 & 0.17 & 0.01 & 3.6 \\ \hline
\end{tabular}  
\\
\begin{tabular}{|c|c|c|c|}
\multicolumn{4}{c}{middle} \\ \hline
~  $x$ ~ & $k_{F0}$ & A & area \\ \hline
1 & 0.30 & 0.053 & 10.6 \\ \hline 
0.93 & 0.29 & 0.031 & 10.1 \\ \hline 
0.88 & 0.28 & 0.004 & 9.5 \\ \hline 
0.76 & 0.25 & 0.03 & 7.5 \\ \hline
\end{tabular}
\\
\begin{tabular}{|c|c|c|c|}
\multicolumn{4}{c}{outer} \\ \hline
~  $x$ ~ & $k_{F0}$ & A & area \\\hline
1 &  0.46 & -0.02 & 24.8  \\ \hline 
0.93 & 0.44 & -0.07 & 23.2  \\ \hline 
0.88 & 0.44 & -0.05 & 22.9 \\ \hline 
0.76 & 0.43 & -0.04 & 22.2 \\ \hline
\end{tabular} } 
\end{center} 
\end{table}

The energy distribution curves (EDCs) measured below $T_c$ and above $T_c$ are shown in Fig. 2 for BaK0.93 and BaK0.88. Figures 2a-2c show the EDCs at $k_F$ of the inner, middle, and outer FSs, respectively, for BaK0.93 ($T_c$ $\sim $ 7 K) measured at 1.5 K (below $T_c$) and 10 K (above $T_c$). Each EDC at $k_F$ is identified with a FS angle $\varphi$ (Fig. 1). The EDCs shown in Fig. 2a-2c were symmetrized with respect to $E_F$, and the results are shown in Fig. 2d-2f, respectively.  For the inner FS, the EDCs at $k_F$ clearly show the superconducting coherence peaks, and the anisotropy of the SC-gap size seem to be rather small. For the middle FS, the line shape of the symmetrized EDCs at $k_F$ is strongly dependent on the FS angle in the vicinity of $E_F$, indicating that the SC gap is highly anisotropic. For the outer FS, the behavior of the symmetrized EDCs are similar to the middle FS. Thus, there is a FS-sheet dependence in the SC-gap sizes of BaK0.93 as in K122. However, a significant difference exists for the outer FS, where the SC-gap size is almost zero for all FS angles for the case of K122. 
Figures 2g-2i show the EDCs at $k_F$ of the inner, middle, and outer FSs, respectively, for BaK0.88 ($T_c$ $\sim $ 13 K) measured at 1.5 K (below $T_c$) and 16 K (above $T_c $). The symmetrized EDCs are shown in Figs. 2g-2i. The anisotropy of the SC-gap size in the middle FS is different from those of K122 and BaK0.93, where the strong anisotropy exists, but similar to that of the OP BaK122, where the anisotropy is weak. For the outer FS, the symmetrized EDCs in Fig. 2l indicate that the SC gap has a strong anisotropy and nodes may exist. This sheet dependence is also different from the other doping levels, OP BaK122, BaK0.93, and K122. The results for BaK0.76 are shown in Fig. S6, and the overall FS-sheet dependence and anisotropy of the SC gap are similar to those of BaK0.88.
 
\begin{figure}[t]
\begin{center}
\includegraphics[width=7.5cm]{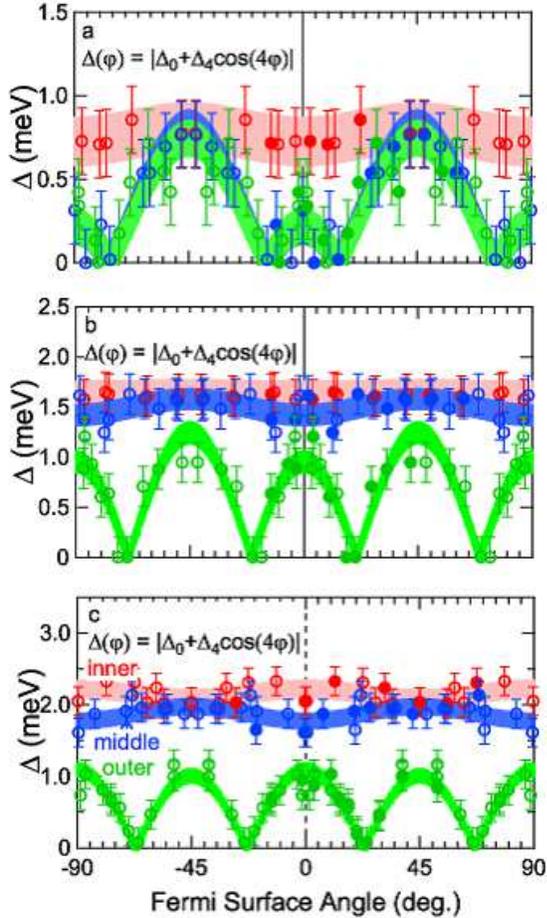}
\caption{Anisotropy of SC gaps and fitting results for each FS. (a)-(c), SC-gap sizes derived from the fitting using a BCS spectral function plotted as a function of FS angle for BaK0.93 (a), BaK0.88 (b), and BaK0.76 (c), respectively. The solid circles are derived from the EDCs shown in Fig. 2 and the open circles are plotted by symmetrizing taking into account the tetragonal crystal symmetry. The error bar for the plots denotes the standard deviation of the $E_F$ position. The thick lines are the fitting functions using a model gap function of $\Delta(\varphi)=|\Delta_0+\Delta_4\cos(4\varphi)|$. The line width corresponds to the standard deviation of the fitting function.}
\end{center}
\end{figure}

In order to quantify the SC-gap size, we have fitted the EDCs to the BCS spectral function.~\cite{Okazaki2012Science} All the EDCs at 1.5 K could be fitted to the calculated spectra, as shown by the solid lines in Fig. 2. We plot the SC-gap sizes derived from fitting as a function of FS angle $\varphi$ in Fig. 3 for BaK0.93 (Fig.3a), BaK0.88 (Fig.3b), and BaK0.76 (Fig.3c) as indicated by the solid circles. The open circles are plotted by symmetrizing data, taking into account the tetragonal crystal symmetry. To obtain insight into the SC-gap anisotropy, we have tried to fit the SC gaps with the two lowest harmonics under the fourfold tetragonal symmetry,
\begin{equation}
\Delta(\varphi)=|\Delta_0+\Delta_4\cos(4\varphi)+\Delta_8\cos(8\varphi)|,
\end{equation}
as in the case of K122. Whereas a $\cos(8\varphi)$ term was finite in the case of K122, we confirmed that the $\cos(8\varphi)$ component is unnecessary to reproduce the SC-gap anisotropy of the measured three compounds of BaK0.93, BaK0.88, and BaK0.76. Thus, we omitted the $\cos(8\varphi)$ term. The fitting is successful for these three compounds as shown in Fig. 3. First of all, the SC-gap symmetry of these three compounds should be $A_{1g}$ type ($s$-wave) as in the case of K122, because the inner FS has no SC-gap nodes. In addition, the anisotropy is strongly suppressed in comparison to K122, where the $\cos(8\varphi)$ component can be recognized. As for the middle FS, the SC-gap nodes exist for BaK0.93, as in the case of K122, and they are determined to be located at $\varphi = \pm(16.0 \pm 2.3)^{\circ}$. On the other hand, for the cases of BaK0.88 and BaK0.76, the anisotropy is strongly suppressed. Finally for the outer FS, strong anisotropies exist for these three compounds, in strong contrast to K122 and BaK0.7, where the SC gaps of the outer FS were vanishingly small. The SC-gap nodes also exist in the outer FS for these three compounds and they are determined to be located at $\varphi = \pm(15.2 \pm 2.8)^{\circ}$ for BaK0.93, $\varphi = \pm(22.2 \pm 0.4)^{\circ}$ for BaK0.88, and $\varphi = \pm(20.2 \pm 1.8)^{\circ}$ for BaK0.76, respectively.

\begin{figure}[t]
\begin{center}
\includegraphics[width=7cm]{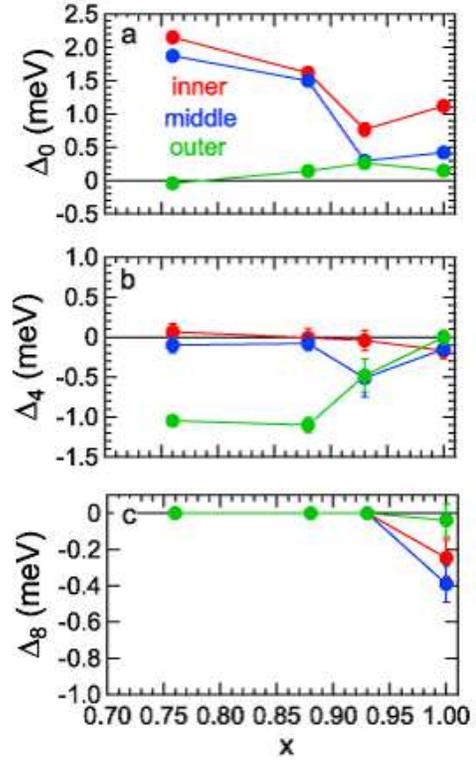}
\caption{Figure 4 $\mid$ Doping dependence of fitting parameters for Fig. 3. (a) isotropic component $\Delta_0$, (b) $\cos(4\varphi)$ component $\Delta_4$, and (c) $\cos(8\varphi)$ component $\Delta_8$, respectively.}
\end{center}
\end{figure}

Figure 4a-4c show the doping dependence of the fitting parameters of $\Delta_0$, $\Delta_4$, and $\Delta_8$, respectively, for the inner, middle, and outer FSs (see Fig. S7 for the ratio to $k_BT_c$, where $k_B$ is the Boltzmann constant).
First, the doping dependence of the isotropic component $\Delta_0$ seems to be not so systematic. Those of the inner and middle FSs are rather small for BaK0.93 compared to the other compounds. That of the outer FS is slightly larger for BaK0.93. As for the $\cos(4\varphi)$ component $\Delta_4$, that of the inner FS is very small for all the compounds. On the other hand, that of the middle FS is large only for the case of BaK0.93. For the case of the outer FS, $\Delta_4$ becomes monotonically larger by Ba doping from K122 to BaK0.88 and it becomes slightly small at BaK0.76. Finally, the $\cos(8\varphi)$ component $\Delta_8$ is finite only for K122. 

At first glance, this doping dependence may seem to be inconsistent with the OP BaK122 and even with OD BaK122, because the OP BaK122 has almost isotropic SC gaps and the SC-gap size of the outer FS is vanishingly small for BaK0.7.~\cite{Malaeb2012PRB} However, if $\Delta_4$ of the outer FS becomes almost zero by further Ba doping, it is completely consistent with BaK0.7. It should be possible because $\Delta_4$ of the middle FS becomes almost zero at BaK0.88 although that of BaK0.93 is rather large. We attributed this doping dependence of SC gap to the existence of competing pairing interactions. Recently, Maiti {\it et al.} proposed that if the intraband and interband interactions of the hole FSs are nearly degenerate, the existence of the SC-gap nodes in the middle FS of K122 can be reproduced.~\cite{Maiti2012PRB} From our results, the components of $\Delta_0$, $\Delta_4$, and $\Delta_8$ change rather drastically with the small amount of Ba doping. This should indicate that two kinds of interactions are nearly degenerate and competing in this doping region, as Maiti {\it et al} proposed that the interband and intraband interactions are nearly degenerate to explain the octet nodal behavior of K122.~\cite{Maiti2012PRB} It can be regarded that because two kinds of interactions are competing in this doping region, the SC-gap anisotropy is very sensitive to the small amount of Ba doping.

If the origin of the competing interactions are interband and intraband interactions as proposed by Maiti {\it et al.}, it is suggested that both of the spin and orbital fluctuations are important. Because the orbital contribution for the outer FS is almost only from the $x^2-y^2$ orbital,~\cite{Shimojima2011Science,Okazaki2012Science} the interband and intraband interactions for this band is inter-orbital and intra-orbital interactions, respectively. In addition, spin and orbital fluctuations can be considered as the origin of these interactions and they are intra-orbital and inter-orbital interactions,~\cite{Kuroki2008PRL,Kontani2010PRL} respectively. Thus, both of the spin and orbital fluctuations seem to be important at the same level at least for the outer FS.

We have investigated the anisotropy of superconducting (SC) gap for several Ba-doped KFe$_2$As$_2$ (K122) samples using laser-based angle-resolved photoemission spectroscopy. A drastic change is observed in the SC-gap anisotropy and nodal behaviors with a small amount of Ba doping. We attribute this drastic doping dependence to the existence of competing pairing interactions in this doping region. From this competition, it is suggested that both of the spin and orbital interactions are important as the origin of the pairing interactions.

We would like to thank Y. Matsuda, T. Shibauchi, H. Kontani, R. Arita, and H. Ikeda for valuable discussions and comments. This research is supported by Grant-in-Aid for JSPS Fellows 23$\cdot$6478, and JSPS through its FIRST Program.

\end{document}